\documentclass[fleqn,twoside]{article}
\usepackage{espcrc2}

\usepackage{epsfig}

\newcommand{\AmS}{{\protect\the\textfont2
  A\kern-.1667em\lower.5ex\hbox{M}\kern-.125emS}}

\title{Looking to the Future: A Fermilab Viewpoint}

\author{H. E. Montgomery\address[MCSD]{Fermi National Accelerator Laboratory, \\
        P.O. Box 500, Batavia, Illinois 60510, U.S.A.}%
}

\begin{document}

\begin{abstract}
This is a short paper summarising a presentation of the evolution
of the Fermilab program for the next five to ten years. Emphasis
is given to the Fermilab accelerator complex, but external
collaboration is emphasised. \vspace{1pc}
\end{abstract}

\maketitle

\section{Introduction}

An informal public discussion was held near the end of the program
of presentations at the High Intensity Frontier Workshop on Elba
in the spring of 2005. This was close to the point (July 1, 2005)
at which Fermilab changed its Director. At the time of the
discussion, the incoming Director, Piermaria Oddone, had made a
presentation\cite{pier-epp2010} giving his vision for the
laboratory. While outlining the current program and looking toward
potential directions for collaboration, this presentation and
paper leans heavily on Oddone's presentation.

\section{Current Program}

The future planning at Fermilab is founded on the current
accelerator-based physics program. This program is dominated by the
Tevatron Collider experiments which enjoy the highest energies
available in the laboratory and will continue to do so until the
Large Hadron Collider (LHC) comes online at CERN later in this
decade. At present approximately one $fb^{-1}$ of integrated
luminosity has been delivered to each experiment. Peak instantaneous
luminosities of greater than $10^{32}~cm^{-2}.~sec^{-1}$ are common
and prospects for yet higher luminosity are bright.

The neutrino program now has several components. The MiniBooNE
experiment is searching for neutrino oscillations corresponding to
the indications observed by the LSND experiment at Los Alamos a few
years ago. This experiment has received approximately $6\times
10^{20}$ protons on target from the Booster Accelerator at 8 GeV; we
are eagerly looking forward to the results. During the past year the
Neutrino at the Main Injector (NuMI) beamline was completed,
commissioned, and is now operating. The MINOS experiment, a long
baseline neutrino oscillation experiment using that beam line, hopes
to receive a good fraction of $10^{20}$ protons on target by years
end. We hope soon to mount the MINER{$\nu$}A neutrino scattering
experiment. To complement these immediate endeavours there has also
been R{\&}D on muon cooling related to future neutrino factories

Fermilab has been a major player in the construction of the
interaction region magnets for the Large Hadron Collider (LHC),
which is being constructed at CERN in Europe. This has been in
collaboration with two other US laboratories and with the KEK
laboratory in Japan. Fermilab is the host laboratory for US-CMS,
the collaboration which is building the U.S. contributions to the
Compact Muon Solenoid (CMS) experiment. Fermilab physicists have
enjoyed leadership r\^oles in the construction project and the
computing and software program. They are now taking the lead in
the maintenance and operation. Recently an LHC Physics Center
(LPC) was initiated; it is hoped that this will enable US
scientists to enhance their places in the analysis phases of the
experiment and that CMS will see increased activity in this vital
area.

The Laboratory is host to a suite of astroparticle physics
experiments which include the Sloan Digital Sky Survey, The Pierre
Auger cosmic ray observatory, and the Cryogenic Dark Matter Search
experiment in the Soudan mine. Each of these is a world leader and
on this base we are developing a new Dark Energy Survey
experiment.

The experimental programs of the laboratory are supported by two
excellent theoretical groups in particle and particle
astrophysics.

Nevertheless, the name of the laboratory contains the word
accelerator. It is in accelerator R\&D for the future, and for the
experiments to go with those accelerators, where we have seen the
largest number of contributions from the laboratory to this
workshop. These are perhaps the most important things we bring to
the table.

\section{New Initiatives}

The overarching goal for the laboratory will be to enable the most
powerful attack possible on the fundamental science questions by
providing world class facilities. These facilities will be
integrated in a global network. It is necessary to develop science
and technology for particle and astroparticle physics research.

Specific goals are to make vital contributions during the next
decade to developing a powerful new tool for discovery, the
International Linear Collider. This will be achieved while
maintaining the foremost neutrino program through the development
of the NO$\nu$A experiment and enhancements to the Fermilab proton
source.

To contribute\cite{mishra-elba2005} to the International Linear
Collider, it is now critical to establish, within the U.S., world
class expertise in super-conducting radio-frequency technology.
There are facilities at Fermilab, buildings which exist, which are
being turned to this goal. The work will be integrated into the
global effort. That effort has an immediate aim to generate a
conceptual design for the machine by the end of 2006. At that
juncture, it is hoped that there will be sufficient information to
inform the choice of paths forward. If the design looks as though it
fits to both the needs of the science and that of the governments
then it will be important to complete the development work to enable
a decision to construct or not as early as possible. If there are
major issues, it may be necessary to embark on a longer development
program.

On the neutrino front, Fermilab is currently operating a neutrino
program which enjoys the most powerful proton beam available to that
end in the world. By year end we can see operation for MINOS the
long baseline oscillation experiment with a beam power of 200 KW.
Upgrades could get that to 450 KW by 2008. If the Tevatron Collider
program ends in 2009, that would make in excess of 600 KW available.
Imaginative ideas suggest potential in the 1 MW range at 120 GeV.

However, the existing complex is old and the path beyond 1 MW with
that complex is daunting. A new source, dubbed the Fermilab Proton
Driver\cite{foster-elba2005} (FPD), would alleviate most of these
issues and provide a platform for high intensity proton source
operation over a range of energies for physics. It is currently
envisaged that this FPD would be an 8 GeV super-conducting linear
accelerator. At low energy, less than 1 GeV, its concepts are
borrowed from the Spallation Neutron Source (SNS) being built at
Oak Ridge National Laboratory. At high energies it would have many
similarities to the Tesla design for the linear collider. There
appears to be considerable synergy with the work needed for the
ILC.

The currently operating MINOS experiment  is well matched by the
current NuMI beam. That beam with the enhancements outlined above
over the next few years could support the search for $\nu_\mu$ to
$\nu_e$ oscillations with a long baseline. This is a key to the
next step in understanding the neutrino flavor physics. The
experiment has been named NO$\nu$A\cite{ray-elba2005}. Its current
design envisages a mass of about 20-30 ktonnes of liquid
scintillator and polyvinyl-chloride tube container. The very long
baseline would enable a unique sensitivity to the neutrino mass
hierarchy. Thus, NO$\nu$A is seen as the second step, after MINOS
in a flexible approach to the neutrino physics puzzles. Since the
figure of merit involves also the detector sensitivity we also
foresee starting a program of R\&D into the use of liquid argon
technology\cite{para-elba2005} which might have potential for the
phase after NO$\nu$A.

The physics that one can achieve with a proton driver is
extensive\cite{geer-elba2005}; it could potentially include high
intensity muon, pion and kaon experiments as demanded by the
science. In fact, when we consider the potential of intense proton
sources for visionary programs such as neutrino factories, the
Fermilab Proton Driver could provide an important and broad launch
platform.

\section{Opportunities for Collaboration}

No single laboratory is going to do all, or even half, of particle
physics alone and Fermilab recognises that collaboration is
essential. This ranges from the International Linear Collider
where Fermilab expects to collaborate on superconducting
technology, in the ILC machine and on the preparations for the
experiments.

The enhancement of the proton source and the R\&D towards a Proton
Driver already takes from the designs of others and even at this
meeting some new opportunities have opened up. It was emphasised
above that a major component of Fermilab research, both
accelerators and experiments is associated with the LHC program at
CERN. We are already involved in R\&D towards a luminosity upgrade
for that machine and we expect this to continue. Rounding out our
accelerator work we have an important but modest involvement in
muon beam cooling and a participation in the MICE experiment.

On the experimental side, we have been delighted by the
participation from all over the world in the Tevatron Collider
program and, to a lesser extent, in the neutrino program. We would
like to see Fermilab continue as one of the homes of the
international particle physics community. To that end, many may be
receiving invitations to join in the NO$\nu$A initiative; we also
expect that the liquid argon initiative will lead to collaboration.

Finally, it is clear that the limited resources have bitten into
the kaon based flavor programs across the world. At Fermilab, the
recent direction has been toward the measurement of the decay
$K\rightarrow{\pi}{\nu}{\overline\nu}$ which appears to be the
most readily accessible. Since it has not been possible to build
such an experiment at Fermilab, we intend to conduct discussions
as to whether a sensible U.S. participation in such an experiment
at CERN could be mounted.

\section{Summary}

At Fermilab, we are enjoying an exciting present. There is a very
healthy program currently operating with luminosity and
intensities increasing regularly and physics results and
publications appearing apace.

For the future we seek to be host to the International Linear
Collider and to build on our neutrino program by improving our
source and preparing the Fermilab Proton Driver.

To achieve this we must expand the extent of our collaboration
with all.

\end{document}